\documentclass[allclo]{FBSart}
\usepackage{epsfig,psfig}
\usepackage[latin1]{inputenc}
\usepackage[T1]{fontenc}
\usepackage{amsmath,amssymb,amsfonts}
\usepackage[rightcaption]{sidecap}
\title{Restoration of chiral symmetry in light-front finite temperature field theory\footnote{Presented by S.~Strauß at Light-Cone 2004, Amsterdam, 16 - 20 August}}
\author{S.~Strauß\footnote{email: {\tt strauss@rubin.physik2.uni-rostock.de}}, M.~Beyer, S.~Mattiello}
\institute{Fachbereich Physik, Universit\"at Rostock, D-18051
  Rostock, Germany}
\runningauthor{S.~Strauß, M.~Beyer, S.~Mattiello}
\runningtitle{Restoration of chiral symmetry in light-front finite temperature field theory}
\newcommand{\ga}{\alpha}
\newcommand{\gb}{\beta}
\newcommand{\gc}{\gamma}
\newcommand{\gd}{\delta}

\newcommand{\gve}{\varepsilon}
\newcommand{\gl}{\lambda}

\newcommand{\kv}{\mbox{\boldmath$k$}}
\newcommand{\qv}{\mbox{\boldmath$q$}}
\begin{document}
\maketitle
\begin{abstract}
We investigate the properties of $qq$ and $q\bar q$ states in hot and
dense quark matter in the framework of light-front finite temperature
field theory. Presently we use the Nambu Jona-Lasinio model of QCD and
derive the gap equation at finite temperature and density. 
We study pionic and scalar diquark dynamics in quark matter 
and calculate the masses and the Mott dissociation as a 
function of the temperature $T$ and the chemical potential $\mu$. 
For the scalar diquark we determine the critical temperature of color
superconductivity.
\end{abstract}
\section{In-medium gap equation in NJL}
Relativistic heavy ion collisions open the possibility to study quantum
chromodynamics (QCD) at finite temperature $T$ and density or chemical
potential $\mu$. This is necessary to understand the evolution of the
early universe and the interior of neutron stars. A way suited to describe the
QCD phases is using light-front quantization. Previously we have used a spinless 
zero-range interaction. Here we use the light-front
form \cite{Dir49} of the Nambu Jona-Lasinio (NJL) model that shows chiral symmetry breaking and
calculate several phase boundaries including the chiral symmetry restoration.
The NJL Lagrangian for the two flavor case reads
\begin{equation}
\label{g1}
{\cal L}_{\rm NJL}=\bar\psi(i\gc_\mu{\partial}^\mu-m_0)\psi+G\left((\bar\psi\psi)^2+(\bar\psi i \gamma_5\mbox{\boldmath{$\tau$}}\psi)^2\right).
\end{equation}
An extensive discussion on the NJL model, the
chiral symmetry breaking, and the properties of light
mesons at finite temperatures and chemical potential in the instant
form can be found in Ref.~\cite{Klevansky:qe}.

On the light-front the in-medium quark propagator~\cite{Beyer:2001bc} is given as
\begin{equation}
\label{g3}
{\cal G}(k)=\left(\gc_\mu{k}_{{\rm on}}^\mu+m\right)\left\{\frac{1-f^+(k^+,\kv_\perp)}{k^2-m^2+i\gve}+\frac{f^+(k^+,\kv_\perp)}{k^2-m^2-i\gve}\right\}
\end{equation}
where $k^2_{\rm{on}}=m^2$ denotes the on-shell 4-momentum. The Fermi functions for a quark $(f^+)$ or antiquark $(f^-)$ are 
\begin{equation}
\label{g4}
f^\pm(k^+,\kv_\perp)=\left[\exp\left\{\frac{1}{T}\left(\frac{1}{2}k^-_{\rm{on}}+\frac{1}{2}k^+\mp\mu\right)\right\}+1\right]^{-1}.
\end{equation}
For the generalization of the gap
equation to finite $T$ and $\mu$ it is most convenient to use the
Dyson equation for the quark propagator and compute the self energy in
Hartree approximation using (\ref{g3}). This leads to the in-medium gap
equation
\begin{equation}
\label{g5}
m(T,\mu)=m_0+24G\int\limits_0^\infty\frac{dk^+}{k^+(2\pi)^3}\int\limits_{\mathbb{R}^2} d^2\kv_\perp m(1-f^+(k^+,\kv_\perp)-f^-(k^+,\kv_\perp)).
\end{equation}

We introduce the Lepage-Brodsky (LB) cut-off scheme to regulate the
divergences in (\ref{g5}) and in the following.

\section{Two-body states}
The two-body bound states are investigated within the Bethe-Salpeter approach. We
are using the following $T$-matrix equation \cite{Ishii:1995bu}
\begin{equation}
\label{g6}
T(k)=K+\int\frac{d^4q}{(2\pi)^4}KS_F(q+k/2)S_F(q-k/2)T(k),
\end{equation}
where $K$ is an appropriate interaction kernel and $S_F(q)$ the dressed
quark propagator. Bound states of mass $M$ are identified with poles
of the $T$-matrix $T(k)$ on the mass-shell $k^2=M^2$.
\subsection{Pion}
The interaction in the pseudoscalar channel is given by
\begin{equation}
\label{g7}
K_{\alpha\beta,\gamma\delta}=-2iG(\gamma_5\tau_i)_{\alpha\beta}(\gamma_5\tau_i)_{\gamma\delta}.
\end{equation}
One introduces a reduced $T$-Matrix $t_\pi(k)$ 
\begin{equation}
\label{g8}
T(k)_{\ga\gb\gc\gd}=(\gc_5\tau_i)_{\ga\gb}t_\pi(k)(\gc_5\tau_i)_{\gc\gd}.
\end{equation}
Inserting the definition (\ref{g8}) into (\ref{g6}) the solution for $t_\pi(k)$ is
obtained 
\begin{equation}
\label{g9}
t_\pi(k)=\frac{-2iG}{1+2G\Pi_\pi(k^2)}
\end{equation}
with $\Pi_\pi(k^2)$ evaluated using (\ref{g3}) for $S_F(k)$
\begin{equation}
\label{g11}
\Pi_\pi(k^2)=-6\int\limits_{LB}\frac{dxd^2\qv_\perp}{x(1-x)(2\pi)^3}\frac{M_{20}^2(x,\qv_\perp)\left(1-f^+(M_{20})-f^-(M_{20})\right)}{M_{20}^2(x,\qv_\perp)-k^2}.
\end{equation}
where $M_{20}^2(x,\qv_\perp)=(\qv_\perp^2+m^2)/x(1-x)$.
For given $T$ and $\mu$ one determines the value of $k^2$ to fulfill the pole condition $1+2G\Pi_\pi(m_\pi^2)=0$.
The results for the pion mass $m_\pi$ are presented in Fig.~1.A.
\begin{figure}[ht]
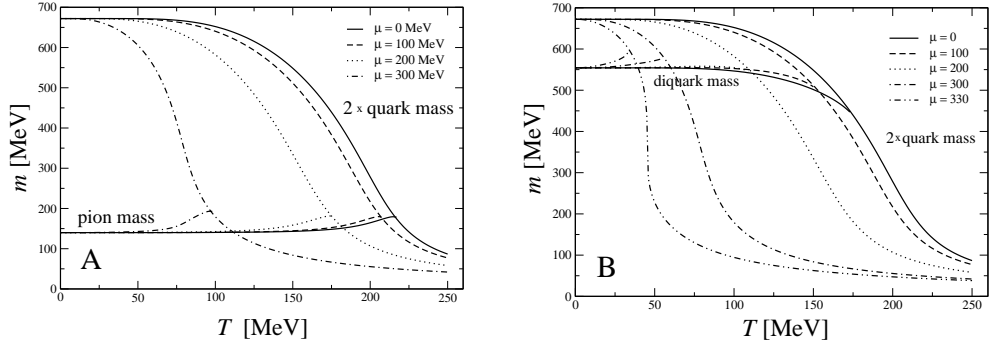

\begin{minipage}[t]{0.48\textwidth}
\centering
\psfig{figure=pionmass.eps,height=45mm}

\end{minipage}
\begin{minipage}{0.04\textwidth}
   \hfill 
\end{minipage}%
\begin{minipage}[t]{0.48\textwidth}
\centering
\psfig{figure=diquark.eps,height=45mm}
\label{fig1}
\end{minipage}
\caption{(A) The pion mass as a function of $T$ for different $\mu$. (B) The diquark mass as a function of $T$ for different $\mu$.
The continuum is given by $2m$. The lines of the two-body masses end at the Mott dissociation points.}
\end{figure}
\subsection{Diquark}
For the scalar, isospin singulet, color-antitriplet diquark the interaction kernel is
\begin{equation}
\label{g12}
K_{\alpha\beta,\gamma\delta}=2iG_s(\gamma_5C\tau_2\gl^a)_{\alpha\beta}(C^{-1}\gamma_5\tau_2\gl^a)_{\gamma\delta}
\end{equation}
with color index $a=2,5,7$ and $C=i\gc^2\gc^0$. 

Following Ref.~\cite{Ishii:1995bu} one can construct the interaction in the diquark
channel by a Fierz transformation of the NJL Lagrangian (\ref{g1}). We treat the
coupling constant $G_S$ as a free parameter to get a reasonably
high diquark mass for $T=\mu=0$. 
The solution of the $T$-matrix equation (\ref{g6}) with the kernel (\ref{g12}) is 
\begin{equation}
\label{g13}
T(k)=(\gc_5C\tau_2\gl^a)_{\ga\gb}\tau_s(k)(C^{-1}\gc_5\tau_2\gl^a)_{\gc\gd}
\end{equation}
with 
\begin{equation}
\label{g14}
\tau_s(k)=\frac{2iG_s}{1+2G_s\Pi_s(k^2)}
\end{equation}  
and 
\begin{equation}
\label{g16}
\Pi_s(k^2)=-6\int\limits_{LB}\frac{dxd^2\qv_\perp}{x(1-x)(2\pi)^3}\frac{M_{20}^2(x,\qv_\perp)\left(1-2f^+(M_{20})\right)}{M_{20}^2(x,\qv_\perp)-k^2}.
\end{equation}
Results for the medium dependence of the diquark mass $m_d$ are shown in Fig.~1.B.

\section{Color superconductivity}
Because of the attractive interaction between quarks one expects, by
Cooper's theorem, a color superconducting phase (CSC) to be present for cold,
dense quark matter. The critical temperature of CSC can be determined 
via the Thouless criterion that 
describes the condensation of the bosonic diquarks, viz. 
\begin{equation}
\label{g17}
m_d(T,\mu)=2\mu.
\end{equation}
Inserting relation (\ref{g17}) the pole condition
$1+2G_S\Pi_s(m_d^2)=0$ for the diquarks becomes
\begin{equation}
\label{g18}
\frac{1}{2G_s}=6\int\limits_{LB}\frac{dx}{x(1-x)}\int\limits_{LB}\frac{d^2\qv_\perp}{(2\pi)^3}\frac{M_{20}^2(x,\qv_\perp)\left(1-2f^+(M_{20})\right)}{M_{20}^2(x,\qv_\perp)-4\mu^2}.
\end{equation}

\begin{SCfigure}[1.0][ht]
\centering
\psfig{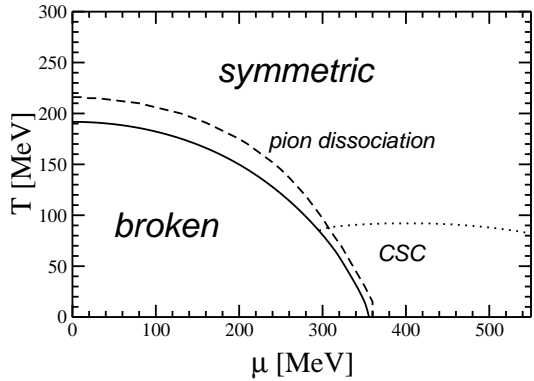}
\caption{The phase diagram in the NJL model. The solid line shows the chiral phase transition. The transition between the deconfined and the CSC phase is given by the dotted line. And the dashed line is the Mott dissociation line of the pion.\vspace{1.7cm}}
\label{fig3}
\end{SCfigure}
Since (\ref{g18}) relates the critical temperature and chemical
potential for CSC one can compute the phase transition from deconfined
to color superconducting quark matter. The result is given in Fig.~\ref{fig3}.

\section{Conclusion}
We have shown that the well known features of the NJL model (chiral
phase transition, diquarks, pion dissociation, superconductivity) are
recovered in a light-front quantization of statistical physics. This
new approach to finite temperature field theory seems to work quite
well. Despite all the challenges that might occur, when applied to
real QCD, this opens a new perspective to explore QCD at finite
temperatures and in particular at larger densities ($\mu>T$), where
lattice QCD is facing severe technical problems.
\section*{Acknowledgment}
We thank the organizers of the Light-Cone 2004 workshop in Amsterdam for this inspiriting meeting.
This work is supported by the Deutsche Forschungsgemeinschaft.


\begin{thebibliography}{99}

\bibitem{Dir49} P.A.M.~Dirac, Rev. Mod. Phys. {\bf 21}, 392 (1949)

\bibitem{Klevansky:qe}
S.~P.~Klevansky, Rev.\ Mod.\ Phys.\  {\bf 64} (1992) 649.
\bibitem{Beyer:2001bc}
M.~Beyer, S.~Mattiello, T.~Frederico and H.~J.~Weber,
Phys.\ Lett.\ B {\bf 521} (2001) 33;
M.~Beyer, S.~Mattiello, T.~Frederico and H.~J.~Weber,
arXiv:hep-ph/0310222.
\bibitem{Ishii:1995bu}
N.~Ishii, W.~Bentz and K.~Yazaki,
Nucl.\ Phys.\ A {\bf 587} (1995) 617;
W.~Bentz, T.~Hama, T.~Matsuki and K.~Yazaki, Nucl.\ Phys.\ A {\bf 651} (1999) 143.
\end{thebibliography}
\end{document}